\documentclass[
    ,final            % use final for the camera ready runs
  ]
  {aipproc}
\layoutstyle{6x9}

\usepackage{graphicx}
\usepackage{amsmath}
\usepackage{color}

\definecolor{darkgreen}{rgb}{0,0.5,0}
\definecolor{purple}{rgb}{0.5,0,0.5}
\definecolor{nblue}{rgb}{0.0,0.0,0.50}
\definecolor{scarlet}{rgb}{1.0,0.2,0}

\newcommand{\lsim}{\mathrel{\rlap{\lower4pt\hbox{\hskip0pt$\sim$}}
\raise1pt\hbox{$<$}}}           %less than or approx. symbol
\newcommand{\gsim}{\mathrel{\rlap{\lower4pt\hbox{\hskip0pt$\sim$}}
\raise1pt\hbox{$>$}}}           %greater than or approx. symbol

\begin{document}

\title{Baryon Properties from Continuum-QCD}


\author{I.\,C.~Clo\"et,$^{1}$
C.\,D.~Roberts$^{2,3}$
and D.\,J.~Wilson$^{2}$}{
  address={
  $^1$Department of Physics, University of Washington, Seattle WA 98195, USA\\
  $^2$Physics Division, Argonne National Laboratory, Argonne, Illinois 60439, USA\\
  $^3$Department of Physics, Institute of High Energy Physics and the State Key Laboratory of Nuclear Physics and Technology, Peking University, Beijing 100871, China
  }}

%\author{Ian C.~Clo\"et}{
%    address={Department of Physics, University of Washington, Seattle WA 98195, USA},
%    %
%    altaddress={Kavli Institute for Theoretical Physics China, CAS, Beijing 100190, China}
%    }

%\author{H.\,L.\,L.~Roberts}{
%    address={Physics Department, University of California, Berkeley, California 94720, USA},
    %
%    altaddress={Physics Division, Argonne National Laboratory, Argonne, Illinois 60439, USA},
    %
%    altaddress={Institut f\"ur Kernphysik, Forschungszentrum J\"ulich, D-52425 J\"ulich, Germany}
%    }

%\author{C.\,D.~Roberts}{
%  address={Physics Division, Argonne National Laboratory, Argonne, Illinois 60439, USA},
  %
%  altaddress={Department of Physics, Center for High Energy Physics and the State Key Laboratory of Nuclear Physics and Technology, Peking University, Beijing 100871, China},
  %
%  altaddress={Kavli Institute for Theoretical Physics China, CAS, Beijing 100190, China},
  %
%  altaddress={Institut f\"ur Kernphysik, Forschungszentrum J\"ulich, D-52425 J\"ulich, Germany}
%  }

\begin{abstract}
We provide an inkling of recent progress in hadron physics made using QCD's Dyson-Schwinger equations, reviewing: the notion of in-hadron condensates and a putative solution of a gross problem with the cosmological constant; a symmetry-preserving computation that simultaneously correlates the masses of meson and baryon ground- and excited-states, and contributes to a resolution of the conundrum of the Roper resonance; and a prediction for the $Q^2$-dependence of $u$- and $d$-quark Dirac and Pauli form factors in the proton, which exposes the critical role played by diquark correlations within the nucleon.
\end{abstract}

\maketitle

With QCD, Nature has given us the sole known example of a strongly-interacting quantum field theory that is defined through degrees-of-freedom which cannot directly be detected.  This empirical fact of \emph{confinement} ensures that QCD is the most interesting and challenging piece of the Standard Model.  It means that building a bridge between QCD and the observed properties of hadrons is one of the key problems for modern science.  In confronting this challenge, steps are being taken with approaches that can rigorously be connected with QCD.  Herein we recapitulate on efforts within the Dyson-Schwinger equation (DSE) framework \cite{Roberts:1994dr,Chang:2010jq}, which provides a continuum perspective on computing hadron properties from QCD.

No solution to QCD will be complete if it does not explain confinement.  This means confinement in the real world, which contains quarks with light current-masses.  That is distinct from the artificial universe of pure-gauge QCD without dynamical quarks, studies of which tend merely to focus on achieving an area law for a Wilson loop and hence are irrelevant to the question of light-quark confinement.  Confinement may be related to the analytic properties of QCD's Schwinger functions \cite{Krein:1990sf} and can therefore be translated into the challenge of charting the infrared behavior of QCD's \emph{universal} $\beta$-function.  This is a well-posed problem whose solution can be addressed in any framework enabling the nonperturbative evaluation of renormalisation constants.

Dynamical chiral symmetry breaking (DCSB); namely, the generation of mass \emph{from nothing}, is a fact in QCD.  This is best seen by solving the DSE for the dressed-quark propagator \cite{Lane:1974he,Bhagwat:2003vw}; i.e., the gap equation.  However, the origin of the interaction strength at infrared momenta, which guarantees DCSB through the gap equation, is unknown.  This relationship ties confinement to DCSB.  The reality of DCSB means that the Higgs mechanism is largely irrelevant to the bulk of normal matter in the universe.  Instead the single most important mass generating mechanism for light-quark hadrons is the strong interaction effect of DCSB; e.g., one can identify it as being responsible for 98\% of a proton's mass.  NB.\ DCSB is also an amplifier of the Higgs' contribution to explicit chiral symmetry breaking \cite{Flambaum:2005kc}.

We note that the hadron spectrum \cite{Holl:2005vu}, and hadron elastic and transition form factors \cite{Cloet:2008re,Aznauryan:2009da} provide unique information about the long-range interaction between light-quarks and, in addition, the distribution of a hadron's characterising properties amongst its QCD constituents.  However, to make full use of extant and forthcoming data, it is necessary to use Poincar\'e covariant theoretical tools that enable the reliable study of hadrons in the mass range $1$-$2\,$GeV.  Crucially, on this domain both confinement and DCSB are germane; and the DSEs provide such a tool.

For the last thirty years, \emph{condensates}; i.e., nonzero vacuum expectation values of local operators, have been used as parameters in order to correlate and estimate essentially nonperturbative strong-interaction matrix elements.  They are also basic to
current algebra analyses.  It is conventionally held that such quark and gluon condensates have a physical existence, which is independent of the hadrons that express QCD's asymptotically realisable degrees-of-freedom; namely, that these condensates
are not merely mass-dimensioned parameters in a theoretical truncation scheme, but in fact describe measurable spacetime-independent configurations of QCD's elementary degrees-of-freedom in a hadronless ground state.

However, it has been argued that this view is erroneous owing to confinement  \cite{Brodsky:2010xf}.  Indeed, it was proven \cite{Maris:1997hd} that the chiral-limit vacuum quark condensate is qualitatively equivalent to the pseudoscalar-meson leptonic decay constant in the sense that both are obtained as the chiral-limit value of well-defined gauge-invariant hadron-to-vacuum transition amplitudes that possess a spectral representation in terms of the current-quark-mass.  Thus, whereas it might sometimes be convenient to imagine otherwise, neither is essentially a constant mass-scale that fills all spacetime.  Hence, in particular, the quark condensate can be understood as a property of hadrons themselves, which is expressed, for example, in their Bethe-Salpeter or light-front wave functions.  In the latter instance, the light-front-instantaneous quark propagator appears to play a crucial role \cite{Brodsky:2010xf,Burkardt:1998dd}

This has enormous implications for the cosmological constant.  The universe is expanding at an ever-increasing rate and theoretical physics has tried to explain this in terms of the energy of quantum processes in vacuum; viz., condensates carry energy and so, if they exist, must contribute to the universe's dark energy, which is expressed in the computed value of the cosmological constant.  The problem is that hitherto all potential sources of this so-called vacuum energy give magnitudes that far exceed the value of the cosmological constant that is empirically determined.  This has been described as ``the biggest embarrassment in theoretical physics'' \cite{Turner:2001yu}.  However, given that, in the presence of confinement, condensates do not leak from within hadrons, then there are no space-time-independent condensates permeating the universe \cite{Brodsky:2010xf}.  This nullifies completely their contribution to the cosmological constant and reduces the mismatch between theory and observation by a factor of $10^{46}$ \cite{Brodsky:2009zd}, and possibly by far more, if technicolour-like theories are the correct paradigm for extending the Standard Model.

In QCD, DCSB is most fundamentally expressed through a strongly momentum-dependent dressed-quark mass \cite{Lane:1974he,Bhagwat:2003vw}; viz., $M(p^2)$ in the quark propagator:
\begin{equation}
S(p) = \frac{1}{i \gamma\cdot p A(p^2) + B(p^2)} = \frac{Z(p^2)}{i \gamma\cdot p + M(p^2)}\,.
\end{equation}
The appearance and behaviour of $M(p^2)$ are essentially quantum field theoretic effects, unrealisable in quantum mechanics.  The running mass connects the infrared and ultraviolet regimes of the theory, and establishes that the constituent-quark and current-quark masses are simply two connected points on a single curve separated by a large momentum interval.  QCD's dressed-quark behaves as a constituent-quark, a current-quark, or something in between, depending on the momentum of the probe which explores the bound-state containing the dressed-quark.  These remarks should make clear that QCD's dressed-quarks are not simply Dirac particles.  This fact has been elucidated further using a novel formulation of the Bethe-Salpeter equation \cite{Chang:2009zb}, most recently in a demonstration that dressed-quarks possess a large, dynamically-generated anomalous chromomagnetic moment, which produces an equally large anomalous electromagnetic moment \cite{Chang:2010hb} that has a material impact on nucleon magnetic form factors \cite{Chang:2011tx} and also very likely on nucleon transition form factors.

Elucidation of the connection between DCSB and dressed-quark anomalous magnetic moments was made possible by the derivation of a novel form for the relativistic bound-state equation \cite{Chang:2009zb}, which is valid and tractable when the quark-gluon vertex is fully dressed.  This has also enabled an exposition of the impact of DCSB on the hadron spectrum.  For example, spin-orbit splitting between ground-state mesons is dramatically enhanced and this is the mechanism responsible for a magnified mass-separation between parity partners; namely, essentially-nonperturbative DCSB corrections to the rainbow-ladder truncation\footnote{This is the leading-order in a nonperturbative, systematic and symmetry-preserving DSE truncation scheme \protect\cite{Munczek:1994zz}.} largely-cancel in the pseudoscalar and vector channels \cite{Bhagwat:2004hn} but add constructively in the scalar and axial-vector channels \cite{Chang:2010jq}.

These facts have been used in a computation of the light-hadron spectrum that simultaneously correlates the masses of meson and baryon ground- and excited-states within a single framework \cite{Roberts:2011cf}.  At the core of the analysis is a symmetry-preserving treatment of a vector-vector contact interaction.  In comparison with relevant quantities the root-mean-square-relative-error/degree-of-freedom is 13\%.

\begin{figure}[t]
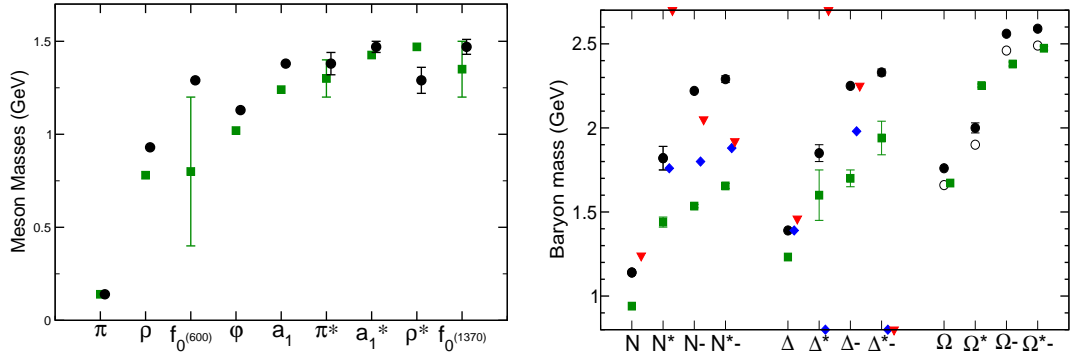

\includegraphics[clip,width=0.45\textwidth]{MesonSpectrum.eps}
\hspace*{1em}\includegraphics[clip,width=0.47\textwidth,height=0.21\textheight]{Spectrum.eps}
\caption{\label{Fig1}
{\sc Left panel} -- Comparison between DSE-computed dressed-quark-core-meson masses and experiment (\emph{filled-squares} \protect\cite{Nakamura:2010zzi}).
{\sc Right panel} -- Comparison between DSE-computed dressed-quark-core-baryon masses (\emph{filled circles}) and bare masses determined in Ref.\,\protect\cite{Suzuki:2009nj} (\emph{filled diamonds}) and Ref.\,\protect\cite{Gasparyan:2003fp} (\emph{filled triangles}).
For the coupled-channels models a symbol at the lower extremity indicates that no associated state is found in the analysis, whilst a symbol at the upper extremity indicates that the analysis reports a dynamically-generated resonance with no corresponding bare-baryon state.
In connection with $\Omega$-baryons the \emph{open-circles} represent a shift downward in the computed results by $100\,$MeV.  This is an estimate of the effect produced by pseudoscalar-meson loop corrections in $\Delta$-like systems at a $s$-quark current-mass \protect\cite{Leinweber:2001ac,Young:2002cj,Eichmann:2008ae}.
}
\end{figure}

The spectra are displayed in Fig.\,\ref{Fig1}.  Notably, with the exception of the $\rho^\ast$-meson, all dressed-quark-core masses lie above the experimental values.  This is readily understood given that the dressed-quark-core is associated with Bethe-Salpeter- and Faddeev-equation kernels that omit contributions which may be associated with pseudoscalar-meson loops.

Following this line of reasoning, a striking feature of the baryon spectrum is agreement between the DSE-computed baryon masses and the bare masses employed in modern dynamical coupled-channels models of pion-nucleon reactions, where the latter exist.  Most interestingly, perhaps, is the Roper resonance.  The DSE study \cite{Roberts:2011cf} produces a radial excitation of the nucleon at $1.82\pm0.07\,$GeV.  This state is predominantly a radial excitation of the quark-diquark system, with both the scalar- and axial-vector diquark correlations in their ground state.  Its predicted mass lies precisely at the value determined in the analysis of Ref.\,\cite{Suzuki:2009nj}.  This is significant because for almost 50 years the ``Roper resonance'' has defied understanding.  Discovered in 1963, it appears to be an exact copy of the proton except that its mass is 50\% greater.  The mass was the problem: hitherto it could not be explained by any theoretical tool that can validly be used to study the strong-interaction piece of the Standard Model of Particle Physics.  That has now changed.  Combined, Refs.\,\cite{Roberts:2011cf,Suzuki:2009nj} demonstrate that the Roper resonance is indeed the proton's first radial excitation, and that its mass is far lighter than normal for such an excitation because the Roper obscures its dressed-quark-core with a dense cloud of pions and other mesons.

Reference~\cite{Roberts:2011cf} provides insight into numerous additional aspects of baryon structure.  For example, relationships between the nucleon and $\Delta$ masses and those of the dressed-quark and diquark correlations they contain, such as $m_N \approx 3\,M$ on a large domain of current-quark mass, where $M$ is the dressed-quark mass; elucidation of a remarkably simple structure for the $\Delta$-resonance; and information on the composition of the excited states of the nucleon and $\Delta$, such as that explained above for the Roper resonance.  It is also a first step in a larger programme aimed at charting the interaction between light-quarks by explicating the impact of differing behaviours of the Bethe-Salpeter kernel upon hadron elastic and transition form factors \cite{Aznauryan:2009da}.

A second step is made in Ref.\,\cite{Roberts:2011wy}, which presents a unified DSE treatment of static and electromagnetic properties of pseudoscalar and vector mesons, and scalar and axial-vector diquark correlations, based upon the same symmetry-preserving treatment of the vector-vector contact-interaction.  This study documents a comparison between the electromagnetic form factors of mesons and those diquarks which play a material role in nucleon structure.  It is therefore an important advance toward a unified description of meson and baryon form factors based on a single interaction.  One notable result is the large degree of similarity between related meson and diquark form factors; e.g., it would be a good practical approximation to assume equality of related radii: $r_{0^+}\approx r_\pi$ and $r_{1^+} \approx r_\rho$, where $r_{0,1}$ are the radii of the scalar and pseudovector diquarks.  This emphasises in addition that the diquark correlations which are important in baryon structure are not pointlike.  A merit of the interaction's simplicity is the ability therewith to compute form factors at arbitrarily large spacelike $Q^2$.  This enabled a zero to be exposed in the $\rho$-meson electric form factor at $z_Q^\rho \approx \surd 6 m_\rho$.  Notably, $r_\rho z_Q^\rho \approx r_{\rm D} z_Q^{\rm D}$, where $r_{\rm D}$ and $z_Q^{\rm D}$ are, respectively, the deuteron's electric radius and the location of the zero in its electric form factor.

The programme just described should be viewed as complementary to Ref.\,\cite{Cloet:2008re}, which uses Schwinger functions constrained by meson phenomena, and perturbative and nonperturbative QCD in order to produce a comprehensive, Poincar\'e-covariant survey of nucleon elastic electromagnetic form factors.  In that body of work, the momentum-dependence of the quark mass-function \cite{Bhagwat:2003vw} plays a critical role; e.g., it has a marked influence on the $Q^2$-evolution of the form factors.
It is through comparison between these results and those anticipated to follow from Refs.\,\cite{Roberts:2011cf,Roberts:2011wy} that one may identify unambiguous signals for the nonperturbative running of the dressed-quark mass-function and the precise nature of the evolution of $M(p^2)$ into the perturbative domain.  Such information is key to charting the long-range behaviour of the quark-quark interaction.

\begin{figure}[t]
\includegraphics[clip,width=0.48\textwidth]{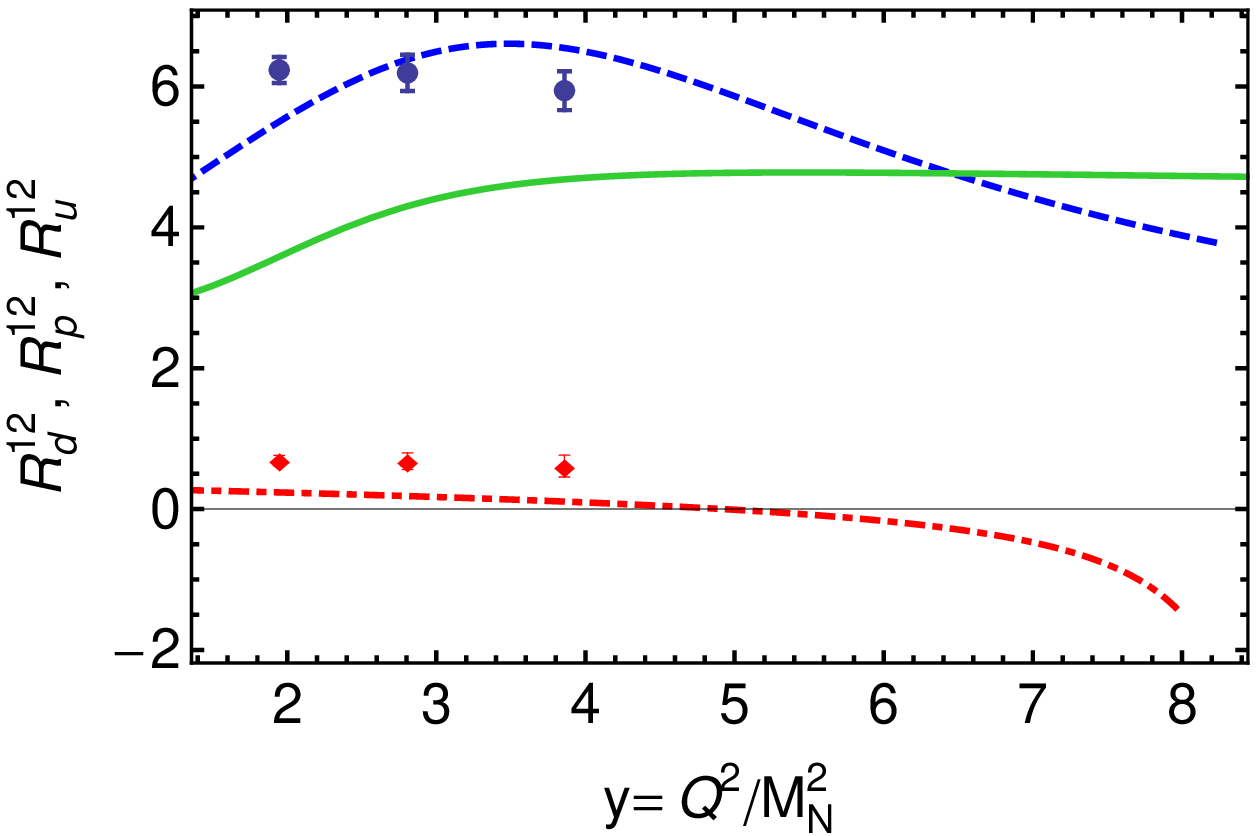}
\hspace*{1em}\includegraphics[clip,width=0.48\textwidth]{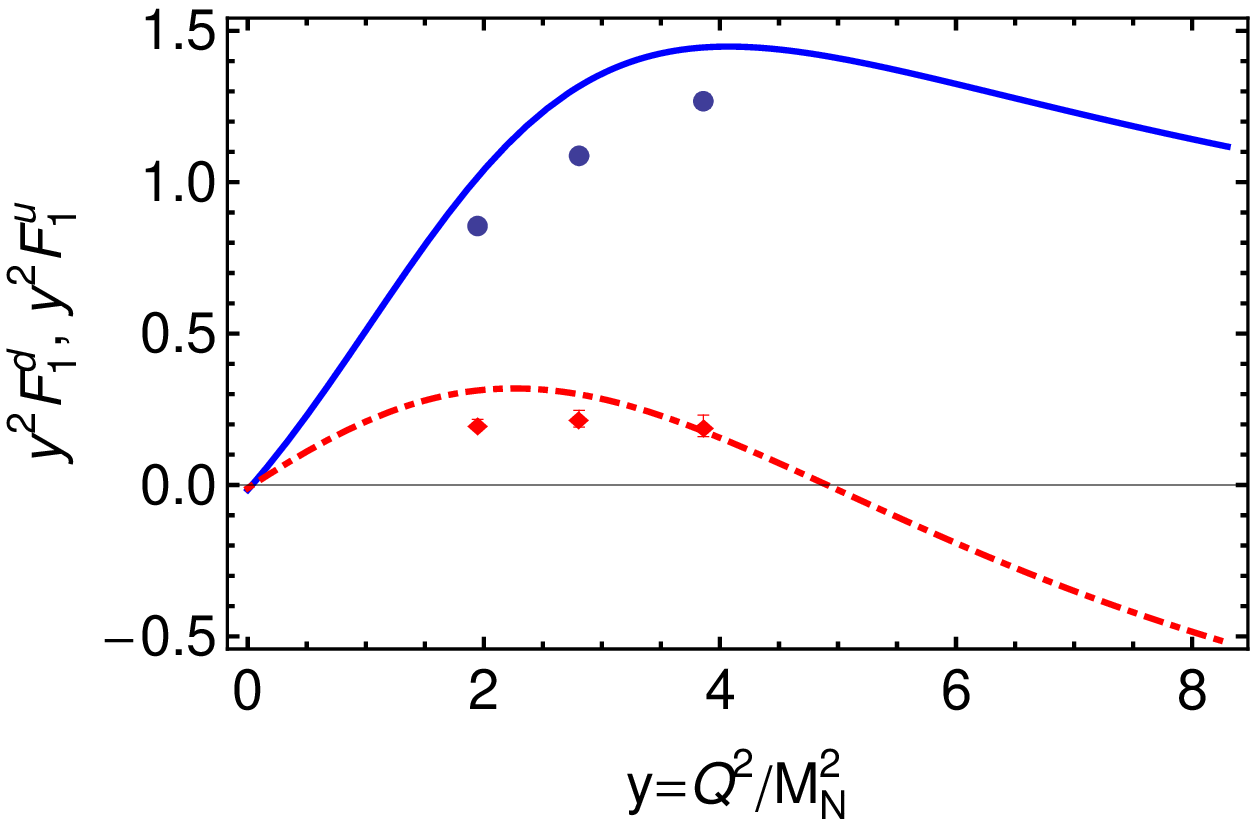}
\caption{\label{Fig2}
{\sc Left panel}.  Ratio defined in Eq.\,(\protect\ref{Rf12}): \emph{dot-dashed curve} -- $d$-quark; \emph{solid curve} -- proton; and \emph{dashed curve} -- $u$-quark.
{\sc Right panel}.  $y^2 F_1^{d,u}(y)$ for light-quarks within the proton: \emph{dot-dashed curve} -- $d$-quark; and \emph{solid curve} -- $u$-quark.
In both panels, the data are drawn from Ref.\,\cite{Cates:2011pz}.
}
\end{figure}

Reference~\cite{Cloet:2008re} predicted the measured behaviour of the neutron's electric form factor \cite{Riordan:2010id}; and it contains a wealth of other information, with much of novelty that remains to be described.  For example in Fig.\,\ref{Fig2}, motivated by Ref.\,\cite{Cates:2011pz}, we plot the ratio
\begin{equation}
\label{Rf12}
R^f_{12}(\hat Q^2) = \frac{(\ln \hat Q^2/\hat \Lambda^2)^2}{\hat Q^2}\,
\frac{\kappa_f\, F_1^f(\hat Q^2)}{F_2^f(\hat Q^2)}\,,
\end{equation}
where $\hat Q^2 = Q^2/M_N^2$, $\hat \Lambda = 0.44$, and $f=$ $u$-quark, $d$-quark or proton.  For the proton, this is the inverse of the ratio depicted in Fig.\,6 of Ref.\,\cite{Cloet:2008re}.  A perturbative analysis that considers effects arising from both the proton's leading- and subleading-twist light-cone wave functions, the latter of which represents quarks with one unit of orbital angular momentum, suggests this ratio should be constant for $Q^2 >\Lambda^2$, where $\Lambda$ is a mass-scale that is supposed to correspond to an upper-bound on the domain of soft momenta \cite{Belitsky:2002kj}.

As explained in Ref.\,\cite{Cloet:2008re} and evident here, $R^p_{12}(\hat Q^2)$ is constant on a sizeable domain, all of which will soon be experimentally accessible.  In this connection, however, the comparison with $R^u_{12}$ and $R^d_{12}$ is revealing: $R^p_{12}(\hat Q^2)$ is constructed from this pair and yet neither is even approximately constant.  This would necessarily have been otherwise if the reasoning of Ref.\,\cite{Belitsky:2002kj} were applicable.  The result $R^p_{12}\approx\,$constant is thus seen to arise only because of cancellations between the separate behaviours of $R^u_{12}$ and $R^d_{12}$, neither of which individually satisfies the scaling relation predicted in Ref.\,\cite{Belitsky:2002kj}.  Hence $R^p_{12}(\hat Q^2)\approx\,$constant is an \emph{accident}.  It is not a verification of the scaling relation but owes to an interference between nonperturbative correlations within the nucleon's Faddeev amplitude.  This explains why, in Fig.\,13 of Ref.\,\cite{Cloet:2008re}, the analogous neutron ratio $R^n_{12}(\hat Q^2)\neq\,$constant: it is constructed from another, distinct combination of $R^u_{12}$ and $R^d_{12}$, and the nature of the interference pattern is therefore different.

These points are amplified by the right panel of Fig.\,\ref{Fig2}, which depicts $Q^4 F_1^d(Q^2)$ and $Q^4 F_1^u(Q^2)$.  Plainly, the scaling behaviour anticipated from perturbative QCD is not evident on the momentum domain depicted.  This fact is emphasised by the zero in $F_1^d(Q^2)$ at $Q^2= 4.9\,M_N^2$, which was first exhibited in Ref.\,\cite{Roberts:2010hu} and is consistent with the extrapolation of an existing empirical form factor parametrisation \cite{Bradford:2006yz}.  The zero owes to the presence of diquark correlations in the nucleon.  It has been found \cite{Cloet:2008re} that the proton's singly-represented $d$-quark is more likely to be struck in association with an axial-vector diquark correlation than with a scalar, and form factor contributions involving an axial-vector diquark are soft.  On the other hand, the doubly-represented $u$-quark is predominantly linked with harder scalar-diquark contributions.  This interference produces a zero in the Dirac form factor of the $d$-quark in the proton.  The location of the zero depends on the relative probability of finding $1^+$ and $0^+$ diquarks in the proton.  The same physics explains the $x = 1$ value of the $d_v(x)/u_v(x)$ ratio of valence-quark distribution functions in the proton \cite{Holt:2010vj}.  Further in this connection, the dressed-quark mass-function underlying these nucleon predictions provides a valid explanation of the valence-quark distributions in pseudoscalar mesons \cite{Hecht:2000xa}, the only one which unifies the phenomena of nonperturbative QCD with the large-$x$ behaviour predicted by QCD.

There are many reasons why this is an exciting time in hadron physics.  We have focused on one.  Namely, through the DSEs, we are positioned to unify phenomena as apparently diverse as: the hadron spectrum; hadron elastic and transition form factors, from small- to large-$Q^2$, and their quark flavour decomposition; and parton distribution functions.  The key is an understanding of both the fundamental origin of nuclear mass and the far-reaching consequences of the mechanism responsible; namely, DCSB.  These things might lead us to an explanation of confinement, the phenomenon that makes nonperturbative QCD the most interesting piece of the Standard Model.

%%%%%%%%%%%%%%%%%%%%%%%%%%%%%%%%%%%%%%%%%%%%
%% MAINMATTER
%%%%%%%%%%%%%%%%%%%%%%%%%%%%%%%%%%%%%%%%%%%%

%%%%%%%%%%%%%%%%%%%%%%%%%%%%%%%%%%%%%%%%%%%%%%%%
%% BACKMATTER
%%%%%%%%%%%%%%%%%%%%%%%%%%%%%%%%%%%%%%%%%%%%%%%%

\bigskip

%\begin{theacknowledgments}
\hspace*{-\parindent}\mbox{\textbf{Acknowledgments.}}~We acknowledge valuable discussions with B.~El-Bennich, V.~Mokeev and B.~Wojtsekhowski.
Work supported by:
the U.\,S.\ Department of Energy, Office of Nuclear Physics, contract nos.~DE-FG03-97ER4014 and DE-AC02-06CH11357.
%\end{theacknowledgments}

\vspace*{-2ex}



\bibliographystyle{aipproc}   % if natbib is available

%\bibliography{sample}

\begin{thebibliography}{99}
%\begin{enumerate}
\bibitem{Roberts:1994dr}
  C.~D.~Roberts and A.~G.~Williams,
  %``Dyson-Schwinger equations and their application to hadronic physics,''
  Prog.\ Part.\ Nucl.\ Phys.\  {\bf 33}, 477 (1994);
%  [arXiv:hep-ph/9403224].
  %%CITATION = PPNPD,33,477;%%
%\bibitem{Roberts:2000aa}
  C.~D.~Roberts and S.~M.~Schmidt,
  %``Dyson-Schwinger equations: Density, temperature and continuum strong
  %QCD,''
  Prog.\ Part.\ Nucl.\ Phys.\  {\bf 45}, S1 (2000);
%  [arXiv:nucl-th/0005064].
  %%CITATION = PPNPD,45,S1;%%
%\bibitem{Maris:2003vk}
  P.~Maris and C.~D.~Roberts,
  %``Dyson-Schwinger equations: A tool for hadron physics,''
  Int.\ J.\ Mod.\ Phys.\  E {\bf 12}, 297 (2003);
%  [arXiv:nucl-th/0301049].
  %%CITATION = IMPAE,E12,297;%%
%\bibitem{Roberts:2007jh}
  C.~D.~Roberts, M.~S.~Bhagwat, A.~H\"oll and S.~V.~Wright,
  %``Aspects of hadron physics,''
  Eur.\ Phys.\ J.\ ST {\bf 140}, 53 (2007).
%  [arXiv:0802.0217 [nucl-th]].
  %%CITATION = 00619,140,53;%%

\bibitem{Chang:2010jq}
  L.~Chang and C.~D.~Roberts,
  ``Hadron Physics: The Essence of Matter,''
  arXiv:1003.5006 [nucl-th].
  %%CITATION = ARXIV:1003.5006;%%

\bibitem{Krein:1990sf}
    C.\,D.~Roberts, A.\,G.~Williams and G. Krein,
  %``On The Implications Of Confinement,''
  Int.\ J.\ Mod.\ Phys.\  A {\bf 7}, 5607 (1992);
  %%CITATION = IMPAE,A7,5607;%%
%\bibitem{Roberts:2007ji}
  C.~D.~Roberts,
  %``Hadron Properties and Dyson-Schwinger Equations,''
  Prog.\ Part.\ Nucl.\ Phys.\  {\bf 61}, 50 (2008).
%  [arXiv:0712.0633 [nucl-th]].
  %%CITATION = PPNPD,61,50;%%

\bibitem{Lane:1974he}
  K.~D.~Lane,
  %``Asymptotic Freedom And Goldstone Realization Of Chiral Symmetry,''
  Phys.\ Rev.\  D {\bf 10}, 2605 (1974);
  %%CITATION = PHRVA,D10,2605;%%
%\bibitem{Politzer:1976tv}
  H.~D.~Politzer,
  %``Effective Quark Masses In The Chiral Limit,''
  Nucl.\ Phys.\  B {\bf 117}, 397 (1976).
  %%CITATION = NUPHA,B117,397;%%

\bibitem{Bhagwat:2003vw}
  M.~S.~Bhagwat, M.~A.~Pichowsky, C.~D.~Roberts and P.~C.~Tandy,
  %``Analysis of a quenched lattice-QCD dressed-quark propagator,''
  Phys.\ Rev.\  C {\bf 68}, 015203 (2003);
%  [arXiv:nucl-th/0304003].
  %%CITATION = PHRVA,C68,015203;%%
%\bibitem{Bhagwat:2006tu}
  M.~S.~Bhagwat and P.~C.~Tandy,
  %``Analysis of full-QCD and quenched-QCD lattice propagators,''
  AIP Conf.\ Proc.\  {\bf 842}, 225 (2006);
%  [arXiv:nucl-th/0601020].
  %%CITATION = APCPC,842,225;%%
%\bibitem{Bhagwat:2007vx}
  M.~S.~Bhagwat, I.~C.~Clo\"et and C.~D.~Roberts,
  ``Covariance, Dynamics and Symmetries, and Hadron Form Factors,''
  arXiv:0710.2059 [nucl-th].
  %%CITATION = ARXIV:0710.2059;%%

\bibitem{Flambaum:2005kc}
  V.~V.~Flambaum \emph{et al}., %A.~Holl, P.~Jaikumar, C.~D.~Roberts and S.~V.~Wright,
  %``Sigma terms of light-quark hadrons,''
  Few Body Syst.\  {\bf 38}, 31 (2006).
%  [arXiv:nucl-th/0510075].
  %%CITATION = FBSYE,38,31;%%

\bibitem{Holl:2005vu}
  A.~H\"oll \emph{et al}., %A.~Krassnigg, P.~Maris, C.~D.~Roberts and S.~V.~Wright,
  %``Electromagnetic properties of ground and excited state pseudoscalar
  %mesons,''
  Phys.\ Rev.\  C {\bf 71}, 065204 (2005).
%  [arXiv:nucl-th/0503043].
  %%CITATION = PHRVA,C71,065204;%%

\bibitem{Cloet:2008re}
  I.~C.~Clo\"et \emph{et al}., %G.~Eichmann, B.~El-Bennich, T.~Klahn and C.~D.~Roberts,
  %``Survey of nucleon electromagnetic form factors,''
  Few Body Syst.\  {\bf 46}, 1 (2009).
%  [arXiv:0812.0416 [nucl-th]].
  %%CITATION = FBSYE,46,1;%%

\bibitem{Aznauryan:2009da}
  I.~Aznauryan {\it et al.},
  ``Theory Support for the Excited Baryon Program at the Jlab 12 GeV Upgrade,''
  arXiv:0907.1901 [nucl-th];
  %%CITATION = ARXIV:0907.1901;%%
%
%\bibitem{Aznauryan:2011ub}
  I.~Aznauryan, V.~D.~Burkert, T.~S.~Lee and V.~Mokeev,
  ``Results from the $N^\ast$ program at JLab,''
  arXiv:1102.0597 [nucl-ex].
  %%CITATION = ARXIV:1102.0597;%%

\bibitem{Brodsky:2010xf}
  S.~J.~Brodsky, C.~D.~Roberts, R.~Shrock and P.~C.~Tandy,
  %``New perspectives on the quark condensate,''
  Phys.\ Rev.\  C {\bf 82}, 022201(R) (2010).
%  [arXiv:1005.4610 [nucl-th]].
  %%CITATION = PHRVA,C82,022201;%%

\bibitem{Maris:1997hd}
  P.~Maris, C.~D.~Roberts and P.~C.~Tandy,
  %``Pion mass and decay constant,''
  Phys.\ Lett.\  B {\bf 420}, 267 (1998);
%  [arXiv:nucl-th/9707003].
  %%CITATION = PHLTA,B420,267;%%
%\bibitem{Maris:1997tm}
  P.~Maris and C.~D.~Roberts,
  %``pi and K meson Bethe-Salpeter amplitudes,''
  Phys.\ Rev.\  C {\bf 56}, 3369 (1997);
%  [arXiv:nucl-th/9708029].
  %%CITATION = PHRVA,C56,3369;%%
%\bibitem{Langfeld:2003ye}
  K.~Langfeld \emph{et al}., %H.~Markum, R.~Pullirsch, C.~D.~Roberts and S.~M.~Schmidt,
  %``Concerning the quark condensate,''
  Phys.\ Rev.\  C {\bf 67}, 065206 (2003).
%  [arXiv:nucl-th/0301024].
  %%CITATION = PHRVA,C67,065206;%%

\bibitem{Burkardt:1998dd}
  M.~Burkardt,
  %``Dynamical vertex mass generation and chiral symmetry breaking on the
  %light-front,''
  Phys.\ Rev.\  D {\bf 58}, 096015 (1998).
%  [arXiv:hep-th/9805088].
  %%CITATION = PHRVA,D58,096015;%%

\bibitem{Turner:2001yu}
  M.~S.~Turner,
  ``Dark energy and the new cosmology,''
  astro-ph/0108103.
  %%CITATION = ASTRO-PH/0108103;%%

\bibitem{Brodsky:2009zd}
  S.~J.~Brodsky and R.~Shrock,
  %``Condensates in Quantum Chromodynamics and the Cosmological Constant,''
  Proc.\ Nat.\ Acad.\ Sci.\  {\bf 108}, 45 (2011).
%  [arXiv:0905.1151 [hep-th]].
  %%CITATION = PNASA,108,45;%%

\bibitem{Chang:2009zb}
  L.~Chang and C.~D.~Roberts,
  %``Sketching the Bethe-Salpeter kernel,''
  Phys.\ Rev.\ Lett.\  {\bf 103}, 081601 (2009).
%  [arXiv:0903.5461 [nucl-th]].
  %%CITATION = PRLTA,103,081601;%%

\bibitem{Chang:2010hb}
L.~Chang, Y.~X.~Liu and C.~D.~Roberts,
  %``Dressed-quark anomalous magnetic moments,''
  Phys.\ Rev.\ Lett.\  {\bf 106}, 072001 (2011).
%  [arXiv:1009.3458 [nucl-th]].
  %%CITATION = PRLTA,106,072001;%%

\bibitem{Chang:2011tx}
  L.~Chang, I.~C.~Clo\"et, C.~D.~Roberts and H.~L.~L.~Roberts,
  ``T(r)opical Dyson-Schwinger Equations,''
  arXiv:1101.3787 [nucl-th].
  %%CITATION = ARXIV:1101.3787;%%

\bibitem{Munczek:1994zz}
  H.~J.~Munczek,
  %``Dynamical chiral symmetry breaking, Goldstone's theorem and the consistency
  %of the Schwinger-Dyson and Bethe-Salpeter Equations,''
  Phys.\ Rev.\  D {\bf 52}, 4736 (1995);
%  [arXiv:hep-th/9411239].
  %%CITATION = PHRVA,D52,4736;%%
%\bibitem{Bender:1996bb}
  A.~Bender, C.~D.~Roberts and L.~Von Smekal,
  %``Goldstone Theorem and Diquark Confinement Beyond Rainbow-Ladder
  %Approximation,''
  Phys.\ Lett.\  B {\bf 380}, 7 (1996).
%  [arXiv:nucl-th/9602012].
  %%CITATION = PHLTA,B380,7;%%

\bibitem{Bhagwat:2004hn}
  M.~S.~Bhagwat \emph{et al}., %A.~H\"oll, A.~Krassnigg, C.~D.~Roberts and P.~C.~Tandy,
  %``Aspects and consequences of a dressed-quark-gluon vertex,''
  Phys.\ Rev.\  C {\bf 70}, 035205 (2004).
%  [arXiv:nucl-th/0403012].
  %%CITATION = PHRVA,C70,035205;%%

\bibitem{Roberts:2011cf}
  H.~L.~L.~Roberts, L.~Chang, I.~C.~Clo\"et and C.~D.~Roberts,
  %``Masses of ground and excited-state hadrons,''
  arXiv:1101.4244 [nucl-th], Few Body Syst.\ \emph{in press}.
  %%CITATION = ARXIV:1101.4244;%%

\bibitem{Nakamura:2010zzi}
  K.~Nakamura {\it et al.}, %  [Particle Data Group],
  %``Review of particle physics,''
  J.\ Phys.\ G {\bf 37}, 075021 (2010).
  %%CITATION = JPHGB,G37,075021;%%

\bibitem{Suzuki:2009nj}
  N.~Suzuki \emph{et al}., %B.~Julia-Diaz, H.~Kamano, T.~S.~Lee, A.~Matsuyama and T.~Sato,
  %``Disentangling the Dynamical Origin of P-11 Nucleon Resonances,''
  Phys.\ Rev.\ Lett.\  {\bf 104}, 042302 (2010).
%  [arXiv:0909.1356 [nucl-th]].
  %%CITATION = PRLTA,104,042302;%%

\bibitem{Gasparyan:2003fp}
  A.~M.~Gasparyan, J.~Haidenbauer, C.~Hanhart and J.~Speth,
  %``Pion nucleon scattering in a meson exchange model,''
  Phys.\ Rev.\  C {\bf 68}, 045207 (2003).
%  [arXiv:nucl-th/0307072].
  %%CITATION = PHRVA,C68,045207;%%

\bibitem{Leinweber:2001ac}
  D.~B.~Leinweber, A.~W.~Thomas, K.~Tsushima and S.~V.~Wright,
  %``Chiral behaviour of the rho meson in lattice QCD,''
  Phys.\ Rev.\  D {\bf 64}, 094502 (2001).
%  [arXiv:hep-lat/0104013].
  %%CITATION = PHRVA,D64,094502;%%

\bibitem{Young:2002cj}
  R.~D.~Young, D.~B.~Leinweber, A.~W.~Thomas and S.~V.~Wright,
  %``Chiral analysis of quenched baryon masses,''
  Phys.\ Rev.\  D {\bf 66}, 094507 (2002).
%  [arXiv:hep-lat/0205017].
  %%CITATION = PHRVA,D66,094507;%%

\bibitem{Eichmann:2008ae}
  G.~Eichmann \emph{et al}., %R.~Alkofer, I.~C.~Cloet, A.~Krassnigg and C.~D.~Roberts,
  %``Perspective on rainbow-ladder truncation,''
  Phys.\ Rev.\  C {\bf 77}, 042202(R) (2008).
%  [arXiv:0802.1948 [nucl-th]].
  %%CITATION = PHRVA,C77,042202;%%

\bibitem{Roberts:2011wy}
  H.~L.~L.~Roberts \emph{et al}., %A.~Bashir, L.~X.~Guti\'errez-Guerrero, C.~D.~Roberts and D.~J.~Wilson,
  ``$\pi$- and $\rho$-mesons, and their diquark partners, from a contact interaction,''
  arXiv:1102.4376 [nucl-th].
  %%CITATION = ARXIV:1102.4376;%%

\bibitem{Riordan:2010id}
  S.~Riordan {\it et al.},
  %``Measurements of the Electric Form Factor of the Neutron up to Q2=3.4 GeV2
  %using the Reaction He3(e,e'n)pp,''
  Phys.\ Rev.\ Lett.\  {\bf 105}, 262302 (2010).
%  [arXiv:1008.1738 [nucl-ex]].
  %%CITATION = PRLTA,105,262302;%%

\bibitem{Cates:2011pz}
  G.~D.~Cates, C.~W.~de Jager, S.~Riordan and B.~Wojtsekhowski,
  ``Flavor decomposition of the elastic nucleon electromagnetic form factors,''
  arXiv:1103.1808 [nucl-ex].
  %%CITATION = ARXIV:1103.1808;%%

\bibitem{Belitsky:2002kj}
  A.~V.~Belitsky, X.~d.~Ji and F.~Yuan,
  %``A perturbative QCD analysis of the nucleon's Pauli form factor  F2(Q**2),''
  Phys.\ Rev.\ Lett.\  {\bf 91}, 092003 (2003).
%  [arXiv:hep-ph/0212351].
  %%CITATION = PRLTA,91,092003;%%

\bibitem{Roberts:2010hu}
  H.~L.~L.~Roberts, L.~Chang, I.~C.~Clo\"et and C.~D.~Roberts,
  ``Exposing the dressed quark's mass,''
  arXiv:1007.3566 [nucl-th].
  %%CITATION = ARXIV:1007.3566;%%

\bibitem{Bradford:2006yz}
  R.~Bradford, A.~Bodek, H.~S.~Budd and J.~Arrington,
  %``A new parameterization of the nucleon elastic form factors,''
  Nucl.\ Phys.\ Proc.\ Suppl.\  {\bf 159}, 127 (2006).
%  [arXiv:hep-ex/0602017].
  %%CITATION = NUPHZ,159,127;%%

\bibitem{Holt:2010vj}
  R.~J.~Holt and C.~D.~Roberts,
  %``Distribution Functions of the Nucleon and Pion in the Valence Region,''
  Rev.\ Mod.\ Phys.\  {\bf 82}, 2991 (2010).
%  [arXiv:1002.4666 [nucl-th]].
  %%CITATION = RMPHA,82,2991;%%

\bibitem{Hecht:2000xa}
  M.~B.~Hecht, C.~D.~Roberts and S.~M.~Schmidt,
  %``Valence-quark distributions in the pion,''
  Phys.\ Rev.\  C {\bf 63}, 025213 (2001);
%  [arXiv:nucl-th/0008049].
  %%CITATION = PHRVA,C63,025213;%%
  %\bibitem{Nguyen:2011jy}
  T.~Nguyen, A.~Bashir, C.~D.~Roberts and P.~C.~Tandy,
  ``Pion and kaon valence-quark parton distribution functions,''
  arXiv:1102.2448 [nucl-th].
  %%CITATION = ARXIV:1102.2448;%%


\end{thebibliography}

\IfFileExists{\jobname.bbl}{}
 {\typeout{}
  \typeout{******************************************}
  \typeout{** Please run "bibtex \jobname" to optain}
  \typeout{** the bibliography and then re-run LaTeX}
  \typeout{** twice to fix the references!}
  \typeout{******************************************}
  \typeout{}
 }

%%%%%%%%%%%%%%%%%%%%%%%%%%%%%%%%%%%%%%%%%%%
%% The following lines show an example how to produce a bibliography
%% without the help of the BibTeX program. This could be used instead
%% of the above.
%%%%%%%%%%%%%%%%%%%%%%%%%%%%%%%%%%%%%%%%%%%

%\end{enumerate}

\end{document}